\documentclass[twocolumn,prl,showpacs,superscriptaddress]{revtex4}
\usepackage{bm}
\usepackage{amssymb}
\usepackage{amsmath}
\usepackage{graphicx}
\usepackage{xspace}
\usepackage{pslatex}

\usepackage[cp1250]{inputenc}

\newcommand{\Cu}{\ensuremath{\mathrm{Cu}}}
\newcommand{\Ni}{\ensuremath{\mathrm{Ni}}}
\newcommand{\Nb}{\ensuremath{\mathrm{Nb}}}
\newcommand{\Al}{\ensuremath{\mathrm{Al}}}

\renewcommand{\O}{\ensuremath{\mathrm{O}}}

\newcommand{\units}[1]{\ensuremath\,{\rm #1}}

\newcommand{\new}[1]{#1}
\newcommand{\q}[1]{}
\newcommand{\Q}[1]{}
\newcommand{\Sec}[1]{}

\begin{document}

\title{%
  $0$-$\pi$ Josephson tunnel junctions with ferromagnetic barrier
}

\author{M. Weides}
\email{m.weides@fz-juelich.de}
\affiliation{%
  Center of Nanoelectronic Systems for Information Technology (CNI), 
  Research Centre J\"ulich, D-52425 J\"ulich, Germany%
}

\author{M. Kemmler}
\author{E. Goldobin}
\email{gold@uni-tuebingen.de}
\affiliation{%
  Physikalisches Institut -- Experimentalphysik II,
  Universit\"at T\"ubingen,
  Auf der Morgenstelle 14,
  D-72076 T\"ubingen, Germany
}

\author{H. Kohlstedt}
\affiliation{%
  Center of Nanoelectronic Systems for Information Technology (CNI), 
  Research Centre J\"ulich, D-52425 J\"ulich, Germany%
}
\affiliation{%
  Department of Material Science and Engineering and Department of Physics, University of Berkeley, California 94720, USA%
}

\author{R. Waser}
\affiliation{%
  Center of Nanoelectronic Systems for Information Technology (CNI), 
  Research Centre J\"ulich, D-52425 J\"ulich, Germany%
}

\author{D. Koelle}
\author{R. Kleiner}
\affiliation{%
  Physikalisches Institut -- Experimentalphysik II,
  Universit\"at T\"ubingen,
  Auf der Morgenstelle 14,
  D-72076 T\"ubingen, Germany
}

\date{\today}

\begin{abstract}
  We fabricated high quality Nb/Al$_2$O$_3$/Ni$_{0.6}$Cu$_{0.4}$/Nb superconductor-insulator-ferromagnet-superconductor Josephson tunnel junctions. Using a ferromagnetic layer with a step-like thickness, we obtain a $0$-$\pi$ junction, with equal lengths and critical currents of 0 and $\pi$ parts. The ground state of our $330\units{\mu m}$ ($1.3\lambda_J$) long junction corresponds to a spontaneous vortex of supercurrent pinned at the 0-$\pi$ step and carrying  $\sim6.7\%$ of the magnetic flux quantum $\Phi_0$. The dependence of the critical current on the applied magnetic field shows a clear minimum in the vicinity of zero field.
\end{abstract}



\maketitle

\Sec{Introduction}

In his classical paper\cite{Josephson:1962:NewEffect} Brian Josephson predicted that the supercurrent through a Josephson junction (JJ) is given by $I_s=I_c\sin(\mu)$. Here, $\mu$ is the Josephson phase (the difference of phases of the quantum mechanical wave functions describing the superconducting condensate in the electrodes), and $I_c>0$ is the critical current (maximum supercurrent that one can pass through the JJ). When one passes no current ($I_s=0$), the Josephson phase $\mu=0$ corresponds to the minimum of energy (ground state). The solution $\mu=\pi$ corresponds to the energy maximum and is unstable.
Later it was suggested that using a ferromagnetic barrier one can realize JJs where $I_s=-I_c\sin(\mu)=I_c\sin(\mu+\pi)$\cite{Bulaevskii:pi-loop}. Such junctions obviously have $\mu=\pi$ in the ground state and, therefore, are called $\pi$ JJs. The solution $\mu=0$ corresponds to the energy maximum and is unstable.

$\pi$ JJs were recently realized using super\-con\-duc\-tor-fer\-ro\-mag\-net-su\-per\-con\-duc\-tor (SFS)\cite{Ryazanov:2001:SFS-PiJJ,Sellier:2004:SFS:HalfIntShapiro,Blum:2002:IcOscillations,Bauer:2004:SFS-SpontSuperCurrents}, super\-con\-duc\-tor-insulator-fer\-ro\-mag\-net-su\-per\-con\-duc\-tor (SIFS)\cite{Kontos:2002:SIFS-PiJJ} and other\cite{Baselmans:1999:SNS-pi-JJ} technologies. In these junctions the sign of the critical current and, therefore, the phase $\mu$ (0 or $\pi$) in the ground state, depends on the thickness $d_F$ of the ferromagnetic layer and on temperature $T$\cite{Buzdin:2005:Review:SF}. $\pi$ JJs may substantially improve parameters of various classical and quantum electronic circuits\cite{Terzioglu:1998:CJJ-Logic,Ustinov:2003:RSFQ+pi-shifters,Ortlepp:2006:RSFQ-0-pi,Ioffe:1999:sds-waveQubit,Yamashita:2005:pi-qubit:SFS+SIS,Yamashita:2006:pi-qubit:3JJ}. \new{To use $\pi$ JJs not only as a ``phase battery'', but also as an active (switching) element in various circuits it is important to have a rather high characteristic voltage $V_c$ (defined e.g. as $V$ at $I=1.2I_c$) and low damping. For example, for classical single flux quantum logic circuits $V_c$ defines the speed of operation. For qubits the value of a quasi-particle resistance $R_{qp}$ at $V=0$ should be high enough since it defines the decoherence time of the circuits. Both high values of $R_{qp}$ and $V_c$ can be achieved by using tunnel SIFS JJs rather than SFS JJs.} 

The dissipation in SIFS JJs decreases exponentially at low temperatures\cite{Weides:2006:SIFS-HiJcPiJJ}, thus, making SIFS technology an appropriate candidate for creating low decoherence quantum circuits, e.g., $\pi$ qubits.\cite{Ioffe:1999:sds-waveQubit,Yamashita:2005:pi-qubit:SFS+SIS,Yamashita:2006:pi-qubit:3JJ}.


Actually, the most interesting situation is when one half of the JJ ($x<0$) behaves as a 0 JJ, and the other half ($x>0$) as a $\pi$ JJ (a 0-$\pi$ JJ)\cite{Bulaevskii:0-pi-LJJ}: In the symmetric case (equal critical currents and lengths of $0$ and $\pi$ parts) the ground state of such a 0-$\pi$ JJ corresponds to a spontaneously formed vortex of supercurrent circulating around the 0-$\pi$ boundary, generating magnetic flux $|\Phi|\leq\Phi_0/2$ inside the junction\cite{Bulaevskii:0-pi-LJJ}. In a very long JJ with length $L\gg\lambda_J$ (Josephson penetration depth) the vortex has the size $\sim\lambda_J$ and carries the flux $\Phi=\pm\Phi_0/2$ --- the so-called semifluxon\cite{Xu:SF-shape,Goldobin:SF-Shape}. Semifluxons are actively studied during the last years\cite{Kirtley:IcH-PiLJJ,Stefanakis:ZFS/2,Goldobin:SF-ReArrange,Zenchuk:2003:AnalXover,Goldobin:Art-0-pi,Susanto:SF-gamma_c,Goldobin:F-SF,Lazarides:Ic(H):SF-Gen,Kato:1997:QuTunnel0pi0JJ,Koyama:2005:d-dot:Qu,Goldobin:2005:MQC-2SFs}. For $L\lesssim\lambda_J$ the vortex does not ``fit'' into the junction and the flux inside the junction $\Phi\approx\pm\Phi_0L^2/(8\pi\lambda_J^2)\ll\Phi_0/2$ \cite{Kirtley:IcH-PiLJJ}. In any case the ground state is degenerate, i.e. may have positive or negative spontaneously formed fractional flux (clockwise or counterclockwise circulating supercurrent) and can be considered as two states (up and down) of a macroscopic spin.


Before, 0-$\pi$ JJs were realized using d-wave superconductors\cite{VanHarlingen:1995:Review,Tsuei:Review,Chesca:2002:YBCO4Crystal-pi-SQUID,Smilde:ZigzagPRL,Ariando:Zigzag:NCCO}, the semifluxons spontaneously formed at the 0-$\pi$ boundary were observed\cite{Kirtley:SF:T-dep,Sugimoto:TriCrystal:SF,Hilgenkamp:zigzag:SF,Kirtley:2005:AFM-SF}, and $I_c(B)$ with a minimum at an external magnetic field $B=0$ was measured\cite{VanHarlingen:1995:Review,Smilde:ZigzagPRL,Ariando:Zigzag:NCCO}. However, the phase shift of $\pi$ in such structures takes place not inside the barrier, but inside the d-wave superconductor. 0-$\pi$ JJs were also obtained (by chance) using SFS technology\cite{DellaRocca:2005:0-pi-SFS:SF,Frolov:2006:SFS-0-pi}, but such structures are quite difficult to measure because of the extremely small $V_c$. In Ref.~\onlinecite{DellaRocca:2005:0-pi-SFS:SF} the presence of spontaneous fractional flux was detected by an auxiliary SIS JJ coupled with the 0-$\pi$ SFS JJ. In Ref.~\onlinecite{Frolov:2006:SFS-0-pi} the $I_c(B)$ was measured using a SQUID-voltmeter.

In this letter we present the first intentionally made \emph{symmetric} 0-$\pi$ \emph{tunnel} JJ of SIFS type with large $V_c$, making direct transport measurements of $I_c(B)$ feasible. Our JJ has a ground state with macroscopic current circulating around the $330\units{\mu m}$ long structure. Such 0, $\pi$ and 0-$\pi$ JJs open a road to self-biased classical and quantum electronic circuits. Moreover, one can study the physics of semifluxons in 0-$\pi$ JJs and, especially, their quantum behavior.

\newcommand{\FigIcBeforEtch}{filled blue circles\xspace}
\newcommand{\FigIcAfterEtch}{open red stars\xspace}
\newcommand{\FigIcTheory}{continuous line\xspace}

\begin{figure}[tb]
  \includegraphics{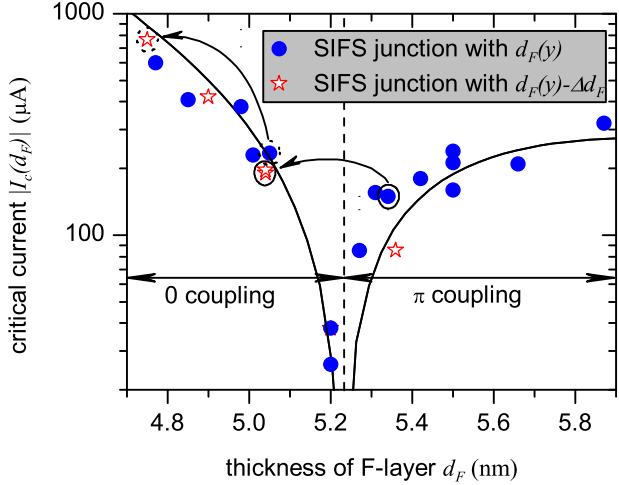}
  \caption{(Color online)
    Dependence of critical current $I_c$ for SIFS reference JJs that were not etched (\FigIcBeforEtch) and etched uniformly (\FigIcAfterEtch) on the thickness of the F-layer $d_F$. Fit of the experimental data for non-etched samples using Eq.~(\ref{Equ:IcRn}) is shown by \FigIcTheory.
  } 
  \label{Fig:Ic(dF)}
\end{figure}

\begin{figure}[tb]
  \begin{center}
    \includegraphics{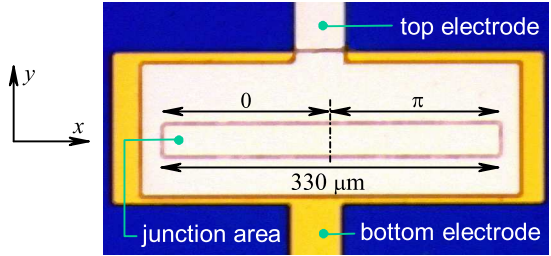}
  \end{center}
  \caption{(Color online)    
    Picture of the $330 \times 30\units{\mu m^2}$ JJs (top view). The 0-$\pi$ boundary/step in the F-layer (if any) is indicated by a dashed line.
  } 
  \label{Fig:sketch}
\end{figure}

The idea of SIFS based 0-$\pi$ JJs is the following. It is known\cite{Buzdin:2005:Review:SF} that the critical current of SIFS JJs changes its sign (and the ground state from $0$ to $\pi$) as a function of the F-layer thickness $d_F$, as shown in Fig.~\ref{Fig:Ic(dF)} by the \FigIcTheory (c.f. Fig.~2 of Ref.~\onlinecite{Kontos:2002:SIFS-PiJJ} or Fig.~1 of Ref.~\onlinecite{Weides:2006:SIFS-HiJcPiJJ}). By choosing two F-layer thicknesses $d_1<d_2$ (e.g. $d_1\approx5\units{nm}$, $d_2\approx5.5\units{nm}$) such that $I_c(d_1)\approx -I_c(d_2)$,  we fabricate a SIFS structure with a step-like $d_F(x)$ to obtain a 0-$\pi$ JJ sketched inside Fig.~\ref{Fig:OurSemifluxon}.

\Sec{Fabrication}

The SIFS junctions were fabricated using a $\Nb/\Al_2\O_3/\Ni_{0.6}\Cu_{0.4}/\Nb$ heterostructure. All 0, $\pi$ and 0-$\pi$ JJs were fabricated within the same technological process. First, Nb-Al$_2$O$_3$ bottom layers were fabricated as for usual high $j_c$ SIS JJs\cite{Weides:SIFS-Tech,Weides:2006:SIFS-HiJcPiJJ}. Second, we have sputtered an F-layer with a gradient of thickness\cite{Weides:SIFS-Tech,Vohl:1991:WedgeF-layer,Kim:2005:Proxy:Nb-Au-CoFe,Zdravkov:2006:SC:Nb-CuNi-bilayers} in $y$ direction. Various structures on the chip where placed within a narrow ribbon along $x$ direction. Such ribbons were replicated along $y$-direction, so that we have the same set of $N$ structures for different $d_F(y)$ along the $y$ axis from $d_F=10\units{nm}$ down to $2.5\units{nm}$ over the 4 in. wafer. After the deposition of a $40\units{nm}$ Nb cap-layer and lift-off we obtain the complete SIFS stack with F-layer thickness $d_F(y)$, but without steps in $d_F$ yet. To produce steps, the (parts of) JJs that are supposed to have larger $d_F$ are protected by photo resist. Then the Nb cap-layer is removed by SF$_6$ reactive rf etching, which leaves a homogeneous flat NiCu surface. About $3\units{\AA}$ ($\Delta d_F$) of NiCu were further Ar ion etched. The above two-step etching and subsequent deposition of a new $40\units{nm}$ Nb cap-layer were done in-situ. \new{Subsequently the junctions were patterned by a three level photolithographic procedure \cite{Gurvitch:1983:NbTech} and insulated by $\Nb_2\O_5$ formed by anodic oxidation after ion-beam etching down to the bottom $\Nb$-electrode.}

Each set of JJs (along one ribbon) has JJs of three classes: (a) not affected by etching with F-layer thickness $d_F(y)$, (b) etched uniformly with F-layer thickness $d_F(y)-\Delta d_F$, and (c) etched to have a step-like $d_F(x)$. All junctions had an area of $10^4\:\mathrm{\mu m^2}$ and lateral sizes comparable to or smaller than $\lambda_J$.

We have measured the critical currents $I_c$ of class (a) JJs (\FigIcBeforEtch) and class (b) JJs (\FigIcAfterEtch) with dimensions $100 \times 100\units{\mu m^2}$ (Fig.~\ref{Fig:Ic(dF)}).

For low-transparency SIFS junctions $I_c(d_F)$ is given by
\begin{equation}
  I_{c}(d_F) \sim
  \exp\left( {-d_F}/{\xi_{F1}} \right)
  \cos \left( (d_F-d_F^\mathrm{dead})/{\xi_{F2}} \right)
  ,\label{Equ:IcRn}
\end{equation}
where $\xi_{F1,F2}$ are the decay and oscillation lengths\cite{Weides:2006:SIFS-HiJcPiJJ}. The coupling changes from $0$ to $\pi$ at the crossover thickness $d_{F}^{x} = \frac{\pi}{2}\xi_{F2}$. Fitting $I_c(d_F)$ of the non-etched junctions (class a) using Eq.~(\ref{Equ:IcRn}), we estimate $\xi_{F1}=0.78 \:\mathrm{nm}$ and $\xi_{F2}=1.35 \:\mathrm{nm}$ and $d_F^\mathrm{dead}=3.09\:\mathrm{nm}$, i.e., $d_{F}^{x}= 5.21\:\rm{nm}$\cite{Weides:2006:SIFS-HiJcPiJJ}. The $|I_{c}(d_F)|$ curve given by Eq.~(\ref{Equ:IcRn}) with such values of $\xi_{F1}$ and $\xi_{F2}$ is shown in Fig.~\ref{Fig:Ic(dF)} by a \FigIcTheory. Comparing it with the experimental $I_c(d_F)$ data for the etched samples (class b) we estimate the etched-away F-layer thickness as $\Delta d_F \approx 3\units{\AA}$. The \FigIcAfterEtch in Fig.~\ref{Fig:Ic(dF)} are shown already shifted by this amount.

Now we choose the set of junctions ($y_0$ of the ribbon) which before etching have thickness $d_2=d_F(y_0)$ and critical current $I_c(d_2)<0$ ($\pi$ junction) and after etching have thickness $d_1=d_2-\Delta d_F$ and critical current $I_c(d_1) \approx -I_c(d_2)$ (0 junction). One of the possibilities is to choose the junction set denoted by closed circles around the data points in Fig.~\ref{Fig:Ic(dF)}, i.e. $d_1=5.05\units{nm}$ and $d_2=5.33\units{nm}$.

Further we deal only with three JJs out of the selected set: (1) a reference SIFS 0 JJ with F-layer thickness $d_{1}$ and critical current density  $j_c^0\equiv j_c(d_1)$; (2) a reference SIFS $\pi$ JJ with F-layer thickness $d_{2}$ and critical current density $j_c^\pi\equiv j_c(d_2)$; and (3) a SIFS 0-$\pi$ JJ with thicknesses $d_1$, $d_2$ and critical current densities $j_c^0$, $j_c^\pi$ in the 0 and $\pi$ halves, respectively. All these junctions have dimensions
$L \times w = 330 \times 33 \units{\mu m^2}$ ($L \parallel x$, $w \parallel y$). The 0-$\pi$ JJ (see Fig.~\ref{Fig:sketch}) consists of one 0 and one $\pi$ region of equal lengths $L_0=L_\pi=165\units{\mu m}$ (within lithographic accuracy).

For all three junctions we have measured the $I$--$V$ characteristics (IVCs) and $I_c(B)$ with $B$ applied in $y$ direction. At $T\approx4.0\units{K}$ the IVCs have no hysteresis and the critical currents of the reference 0 and $\pi$ JJs were $I_c^{0}\approx208\units{\mu A}$, $I_c^{\pi}\approx171\units{\mu A}$, respectively. The dependences $I_c^{0}(B)$ and $I_c^{\pi}(B)$ are almost perfect Fraunhofer patterns, shown in Fig.~\ref{Fig:0pi:Ic(B)}(a). For the 0-$\pi$ JJ, $I_c^{0\text{-}\pi}(B)$ is somewhat asymmetric (e.g. near the first minimum) because of  $j_c^0 \neq j_c^\pi$, but has a clear minimum near zero field. To achieve more symmetric configuration we have measured $I_c(B)$ for all three JJs in a temperature range down to $2.3\units{K}$, because decreasing temperature should increase $I_c^\pi=I_c(d_2)$ faster than $I_c^0=I_c(d_1)$. Two effects are responsible for this behavior. First, when $T$ decreases, the 0-$\pi$ crossover thickness $d_F^x(T)$ decreases, decreasing $I_c^0$ and increasing $I_c^\pi$. Second, the whole amplitude of $I_c(d_F)$ grows as the temperature decreases, similar to the Ambegaokar-Bartoff dependence for conventional SIS JJ. For $I_c^0$ these two dependences have opposite effect, while for $I_c^{\pi}$ they add up. Thus, as $T$ decreases, $I_c^{\pi}$ increases faster than $I_c^0$.

\new{While cooling down and making measurements at each $T$, one of the JJs ($0$, $\pi$ or $0$--$\pi$) after $\sim10\,$h was eventually trapping some flux that we associate with rearrangement of the domains in the F-layer --- $I_c(B)$ was suddenly shifting along the $B$-axis. After thermal cycling, the same symmetric $I_c(B)$ could be measured again.}

\newcommand{\FigIcHz}{red filled triangles}
\newcommand{\FigIcHp}{blue open triangles}
\newcommand{\FigIcHzp}{black spheres}

\begin{figure}[!tb]
  \includegraphics{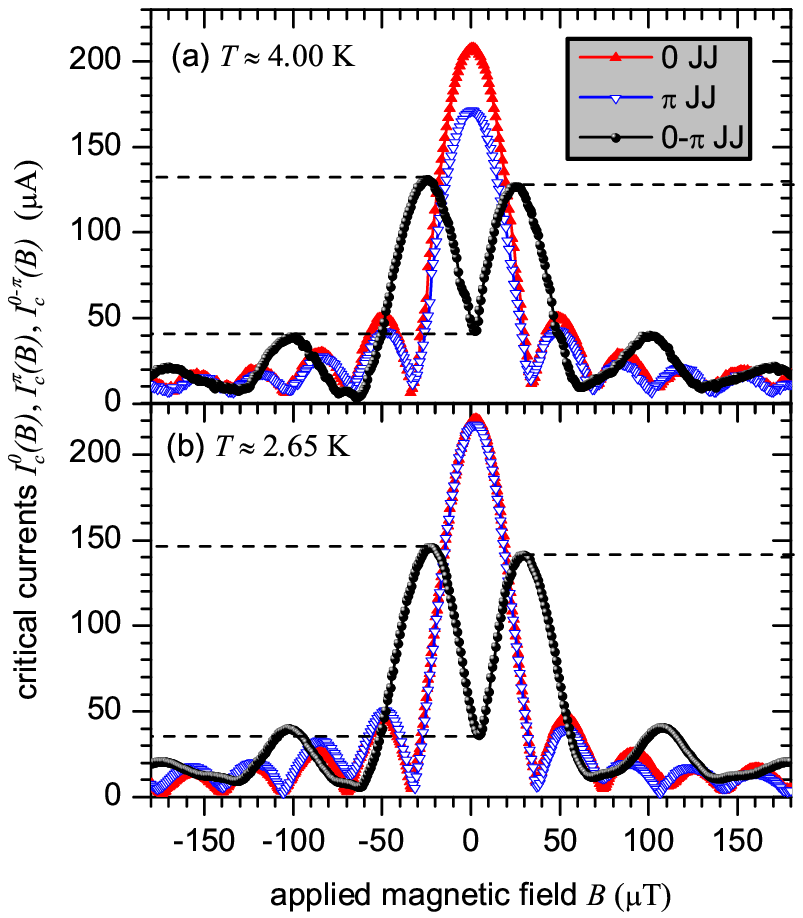}
  \caption{(Color online)
    $I_c(B)$ of $0$ JJ (\FigIcHz), $\pi$ JJ (\FigIcHp) and 0-$\pi$ JJ (\FigIcHzp) measured at (a) $T\approx4.2\units{K}$ and (b) $T \approx 2.65\units{K}$.
  }
  \label{Fig:0pi:Ic(B)}
\end{figure}

The main experimental result of the paper is presented in Fig.~\ref{Fig:0pi:Ic(B)}(b), which shows $I_c(B)$ for all three junctions at $T\approx2.65\units{K}$. At this temperature $I_c^{0}(B)$ and $I_c^{\pi}(B)$  almost coincide, yielding Fraunhofer patterns with critical currents $I_c^{0}\approx 220\units{\mu A}$, $I_c^{\pi}\approx 217\units{\mu A}$ and the same period of modulation.
$I_c^{0\text{-}\pi}(B)$ has a clear minimum near zero field and almost no asymmetry --- the critical currents at the left and right maxima ($146\units{\mu A}$ and $141\units{\mu A}$) differ by less than $4\units{\%}$. 

To ensure that the dip on $I_c^{0\text{-}\pi}(B)$ near zero field originates from $0$-$\pi$ we also measured $I_c^{0\text{-}\pi}(B)$ by applying a field along $x$ direction. In this case the $I_c^{0\text{-}\pi}(B)$ pattern looks like a Fraunhofer pattern with maximum at zero field (not shown).

\Sec{Discussion}

Let us discuss the features of this $I_c^{0\text{-}\pi}(B)$ dependence and its meaning. The Josephson penetration depth 
$
  \lambda_J 
  \approx 259\pm17\units{\mu m}
$ 
is estimated by taking the London penetration depth $\lambda=90\pm10\units{nm}$, the thicknesses of the superconducting electrodes $t_1=120\pm10\units{nm}$ and $t_2=400\units{nm}$, and $j_c=I_c^0/(Lw)\approx2\units{A/cm^2}$. 
%
Thus, the normalized length of our JJs is $l=L/\lambda_J\approx1.3$ at $T=2.65\units{K}$. 

\begin{figure}[!tb]
  \includegraphics{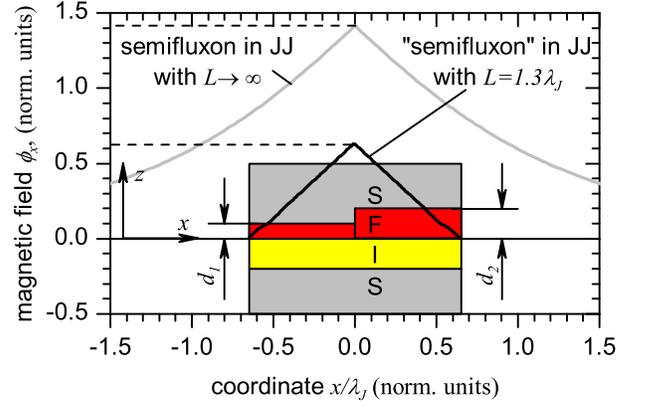}
  \caption{(Color online)
    Numerically calculated magnetic field of spontaneous fractional flux in 0-$\pi$ JJ of $\ell=1.3$. The field of semifluxon in an infinite LJJ is shown for comparison.
  }
  \label{Fig:OurSemifluxon}
\end{figure}

What is the ground state ($I=B=0$) of our 0-$\pi$ JJ? For the symmetric 0-$\pi$ LJJ of length $l=1.3$ the ground state has a spontaneous flux\cite{Kirtley:IcH-PiLJJ} $\pm\Phi=\Phi_0 l^2/8\pi\approx0.067\Phi_0$, i.e. $13\units{\%}$ of $\Phi_0/2$. If the 0-$\pi$ JJ is asymmetric, e.g., $j_c^0 \neq j_c^\pi$, the ground state may correspond to $\mu=0$ or $\mu=\pi$. In our case, using formulas from Ref.~\onlinecite{Bulaevskii:0-pi-LJJ} and our value of $l=1.3$, we estimate that the ground state with spontaneous flux exists at $j_c^\pi/j_c^0$ from $0.78$ to $1.39$. Thus \emph{we are clearly inside the domain with spontaneous flux in the ground state} for $T=2.3\ldots4.2K$, although one cannot see any striking indications of spontaneous flux on $I_c^{0\text{-}\pi}(B)$. Note, that for shorter JJs ($l\lesssim 0.5$) the range of $1-l^2/6<j_c^\pi/j_c^0<1+l^2/6$ with spontaneous flux in the ground state may be extremely small. We would like to point out that even if $j_c^\pi/j_c^0$ is off this domain, e.g., at higher $T$ when the asymmetry is even larger, and the ground state is flat ($\mu=0$ in this case), by applying a bias current (even at $B=0$) or magnetic field one immediately induces fractional flux in the system.\cite{Goldobin:SF-ReArrange} The magnetic field corresponding to our ``semifluxon'' at $T\approx2.65\units{K}$ is shown in Fig.~\ref{Fig:OurSemifluxon}

Using a short JJ model, i.e. assuming that the phase $\mu(x)$ is a linear function of $x$, we can calculate that the first minimum of the Fraunhofer dependence for a 0 or for a $\pi$ JJ should be at $B_{c1} = {\Phi_0}/{L \Lambda}\approx 44.6\pm3.5\units{\mu T}$, 
where the effective magnetic thickness of the junction 
$
  \Lambda 
  \approx 140\pm 11 \units{nm}
$. 
As we see the experimental value of $34\units{\mu T}$ is lower by a factor of $1.3$ due to field focusing. In a short JJ model, $I_c^{0\text{-}\pi}(B)$  should look like\cite{VanHarlingen:1995:Review}
\begin{equation}
  I_c^{0\text{-}\pi}(B) = I_c^0 \frac{\sin^2(\pi f/2)}{|\pi f/2|}
  , \label{Eq:0-pi:Ic(B)}
\end{equation}
where $f=\Phi_\Sigma/\Phi_0=BL\Lambda/\Phi_0$ is the applied number of flux quanta through the effective junction area $L\Lambda$ ($\Lambda\approx2\lambda$). This dependence has $I_c^{0\text{-}\pi}(0)=0$, two symmetric maxima at $I_c^{0\text{-}\pi}(B_m)/I_c^0\approx0.72$ and the first side minima at $f=\pm2$, which should have a parabolic shape touching the $B$ axis. We have some discrepancies between the simple short junction theory (\ref{Eq:0-pi:Ic(B)}) and experiment (Fig.~\ref{Fig:0pi:Ic(B)}). 

First, in our experiment the minimum of $I_c^{0\text{-}\pi}(B)$ is somewhat lifted from zero up to $I_c^{0\text{-}\pi}/I_c^0\approx0.16$. Second, the critical current at the side maxima $I_c^{0\text{-}\pi}(-B_m)/I_c^0\approx0.66$, $I_c^{0\text{-}\pi}(+B_m)/I_c^0\approx0.64$ are below the theoretical value of 0.72 and are a little bit different. Third, the first side minima of $I_c^{0\text{-}\pi}(B)$ are reached at the same $B$ as the second minima of $I_c(B)$ for the 0 or $\pi$ JJs, which is good, but the minima look oblate-shaped from the bottom and do not reach zero.

All these effects can be explained and reproduced numerically by taking into account several additional ingredients. First, the value of $I_c^{0\text{-}\pi}(B)$ at the central minimum is affected by the finite length of the junction, i.e.\@ the deviation from the short JJ model increases $I_c^{0\text{-}\pi}(B)$ at the central minimum. Second, if we include $j_c^\pi=j_c^0(1-2\delta) \neq j_c^0$ in the short JJ model, instead of curve (\ref{Eq:0-pi:Ic(B)}), we will get a \emph{symmetric} curve with $I_c^{0\text{-}\pi}(0)/I_c^0=\delta$, and maximum $I_c^{0\text{-}\pi}(\pm B_m)$ below 0.72. This explains why in Fig.~\ref{Fig:0pi:Ic(B)}(a) the value of $I_c^{0\text{-}\pi}(B)$ at the minimum is larger than in Fig.~\ref{Fig:0pi:Ic(B)}(b). If, instead we assume some weak net magnetization of the F-layer, such that $M_0$ (in the 0 part) is not equal to $M_\pi$ (in the $\pi$ part), we find that $I_c^{0\text{-}\pi}(B)$ shifts along the $B$ axis by $(M_0+M_\pi)/2$, as seen in Fig.~\ref{Fig:0pi:Ic(B)}. If we include \emph{both} assumptions, $I_c^{0\text{-}\pi}(B)$ will also get asymmetric maxima and the characteristic oblate-shape at the first side minima. Details will be presented elsewhere.


\Sec{Conclusions}

In summary, we have fabricated high quality $\Nb/\Al_2\O_3/\Ni_{0.6}\Cu_{0.4}/\Nb$ superconductor-insulator-ferromagnet-superconductor Josephson tunnel junctions. Using a ferromagnetic layer with a step-like thickness, we obtained a $0$-$\pi$ junction, which becomes symmetric at $T\approx2.65\units{K}$ and carries spontaneous fractional flux $\sim0.067\Phi_0$ in the ground state. The dependence of critical current on the applied magnetic field shows a clear minimum in the vicinity of zero field and is well described by Eq.~(\ref{Eq:0-pi:Ic(B)}). \new{In essence our $I_c(B)$ data are the same as for 0-$\pi$ JJs based of d-wave superconductors such as corner JJs\cite{VanHarlingen:1995:Review}, bi-, tri- and tetra- crystal JJs\cite{Tsuei:Review,Chesca:2002:YBCO4Crystal-pi-SQUID}, and YBCO-Nb or NCCO-Nb ramp zigzag JJs\cite{Smilde:ZigzagPRL,Ariando:Zigzag:NCCO}. This is not suprising since the underlaying model\cite{Xu:SF-shape,Kirtley:IcH-PiLJJ,Goldobin:SF-Shape} is the same.} To our knowledge our SIFS 0-$\pi$ JJ is the first underdamped \emph{tunnel} 0-$\pi$ junction based on low-$T_c$ superconductors. It can be measured using standard setups due to the rather high characteristic voltage $V_c$. The possibility to fabricate 0, $\pi$ and 0-$\pi$ Josephson junctions within the same process, having the same $I_c$ and $V_c$ opens perspectives for application of SIFS technology in complimentary logic circuits\cite{Terzioglu:1998:CJJ-Logic}, in RSFQ with active $\pi$ junctions\cite{Ortlepp:2006:RSFQ-0-pi}, in $\pi$ qubits \cite{Ioffe:1999:sds-waveQubit,Yamashita:2005:pi-qubit:SFS+SIS,Yamashita:2006:pi-qubit:3JJ} as well as for the investigation of semifluxons. Due to exponentially decreasing damping for $T\to 0$\cite{Weides:2006:SIFS-HiJcPiJJ} \new{(c.f. in d-wave based 0-$\pi$ JJs damping is larger and does not decrease exponentially)}, SIFS 0-$\pi$ JJs are promising devices for observation of macroscopic quantum effects using semifluxons (macroscopic spins) and for qubits\cite{Kato:1997:QuTunnel0pi0JJ,Goldobin:2005:MQC-2SFs}.

We thank A.~Buzdin, V.~ Oboznov and V.~Ryzanov for useful discussions. This work is supported by the ESF program PiShift and by the Deutsche Forschungsgemeinschaft projects GO-1106/1, KL-930/10 and SFB/TR 21. E.G. acknowledges support by the Elitef\"orderprogram of the Landesstiftung Baden-W\"urttemberg.

\bibliographystyle{apsprl}
\bibliography{this,JJ,SFS,SF,pi,QuComp}%

\begin{thebibliography}{46}
\expandafter\ifx\csname natexlab\endcsname\relax\def\natexlab#1{#1}\fi
\expandafter\ifx\csname bibnamefont\endcsname\relax
  \def\bibnamefont#1{#1}\fi
\expandafter\ifx\csname bibfnamefont\endcsname\relax
  \def\bibfnamefont#1{#1}\fi
\expandafter\ifx\csname citenamefont\endcsname\relax
  \def\citenamefont#1{#1}\fi
\expandafter\ifx\csname url\endcsname\relax
  \def\url#1{\texttt{#1}}\fi
\expandafter\ifx\csname urlprefix\endcsname\relax\def\urlprefix{URL }\fi
\providecommand{\bibinfo}[2]{#2}
\providecommand{\eprint}[2][]{\url{#2}}

\bibitem[{\citenamefont{Josephson}(1962)}]{Josephson:1962:NewEffect}
\bibinfo{author}{\bibfnamefont{B.~D.} \bibnamefont{Josephson}},
  \bibinfo{journal}{Phys. Lett.} \textbf{\bibinfo{volume}{1}},
  \bibinfo{pages}{251} (\bibinfo{year}{1962}).

\bibitem[{\citenamefont{Bulaevski\u{i} et~al.}(1977)}]{Bulaevskii:pi-loop}
\bibinfo{author}{\bibfnamefont{L.~N.} \bibnamefont{Bulaevski\u{i}}},
  \bibnamefont{et~al.}, \bibinfo{journal}{JETP Lett.}
  \textbf{\bibinfo{volume}{25}}, \bibinfo{pages}{290} (\bibinfo{year}{1977}).

\bibitem[{\citenamefont{Ryazanov et~al.}(2001)}]{Ryazanov:2001:SFS-PiJJ}
\bibinfo{author}{\bibfnamefont{V.~V.} \bibnamefont{Ryazanov}},
  \bibnamefont{et~al.}, \bibinfo{journal}{Phys. Rev. Lett.}
  \textbf{\bibinfo{volume}{86}}, \bibinfo{pages}{2427} (\bibinfo{year}{2001}).

\bibitem[{\citenamefont{Sellier
  et~al.}(2004)}]{Sellier:2004:SFS:HalfIntShapiro}
\bibinfo{author}{\bibfnamefont{H.}~\bibnamefont{Sellier}},
  \bibnamefont{et~al.}, \bibinfo{journal}{Phys. Rev. Lett.}
  \textbf{\bibinfo{volume}{92}}, \bibinfo{eid}{257005} (\bibinfo{year}{2004}).

\bibitem[{\citenamefont{Blum et~al.}(2002)}]{Blum:2002:IcOscillations}
\bibinfo{author}{\bibfnamefont{Y.}~\bibnamefont{Blum}}, \bibnamefont{et~al.},
  \bibinfo{journal}{Phys. Rev. Lett.} \textbf{\bibinfo{volume}{89}},
  \bibinfo{pages}{187004} (\bibinfo{year}{2002}).

\bibitem[{\citenamefont{Bauer
  et~al.}(2004)}]{Bauer:2004:SFS-SpontSuperCurrents}
\bibinfo{author}{\bibfnamefont{A.}~\bibnamefont{Bauer}}, \bibnamefont{et~al.},
  \bibinfo{journal}{Phys. Rev. Lett.} \textbf{\bibinfo{volume}{92}},
  \bibinfo{eid}{217001} (\bibinfo{year}{2004}).

\bibitem[{\citenamefont{Kontos et~al.}(2002)}]{Kontos:2002:SIFS-PiJJ}
\bibinfo{author}{\bibfnamefont{T.}~\bibnamefont{Kontos}}, \bibnamefont{et~al.},
  \bibinfo{journal}{Phys. Rev. Lett.} \textbf{\bibinfo{volume}{89}},
  \bibinfo{pages}{137007} (\bibinfo{year}{2002}).

\bibitem[{\citenamefont{Baselmans et~al.}(1999)}]{Baselmans:1999:SNS-pi-JJ}
\bibinfo{author}{\bibfnamefont{J.~J.~A.} \bibnamefont{Baselmans}},
  \bibnamefont{et~al.}, \bibinfo{journal}{Nature}
  \textbf{\bibinfo{volume}{397}}, \bibinfo{pages}{43} (\bibinfo{year}{1999}).

\bibitem[{\citenamefont{Buzdin}(2005)}]{Buzdin:2005:Review:SF}
\bibinfo{author}{\bibfnamefont{A.~I.} \bibnamefont{Buzdin}},
  \bibinfo{journal}{Rev. Mod. Phys.} \textbf{\bibinfo{volume}{77}},
  \bibinfo{eid}{935} (\bibinfo{year}{2005}).

\bibitem[{\citenamefont{Terzioglu and Beasley}(1998)\citenamefont{Terzioglu
  et~al.}}]{Terzioglu:1998:CJJ-Logic}
\bibinfo{author}{\bibfnamefont{E.}~\bibnamefont{Terzioglu}}
  \bibnamefont{et~al.}, \bibinfo{journal}{IEEE Trans. Appl. Supercond.}
  \textbf{\bibinfo{volume}{8}}, \bibinfo{pages}{48} (\bibinfo{year}{1998}).

\bibitem[{\citenamefont{Ustinov and Kaplunenko}(2003)\citenamefont{Ustinov
  et~al.}}]{Ustinov:2003:RSFQ+pi-shifters}
\bibinfo{author}{\bibfnamefont{A.~V.} \bibnamefont{Ustinov}}
  \bibnamefont{et~al.}, \bibinfo{journal}{J. Appl. Phys.}
  \textbf{\bibinfo{volume}{94}}, \bibinfo{pages}{5405} (\bibinfo{year}{2003}).

\bibitem[{\citenamefont{Ortlepp et~al.}(2006)}]{Ortlepp:2006:RSFQ-0-pi}
\bibinfo{author}{\bibfnamefont{T.}~\bibnamefont{Ortlepp}},
  \bibnamefont{et~al.}, \bibinfo{journal}{Science}
  \textbf{\bibinfo{volume}{312}}, \bibinfo{pages}{1495} (\bibinfo{year}{2006}).

\bibitem[{\citenamefont{Ioffe et~al.}(1999)}]{Ioffe:1999:sds-waveQubit}
\bibinfo{author}{\bibfnamefont{L.~B.} \bibnamefont{Ioffe}},
  \bibnamefont{et~al.}, \bibinfo{journal}{Nature (London)}
  \textbf{\bibinfo{volume}{398}}, \bibinfo{pages}{679} (\bibinfo{year}{1999}).

\bibitem[{\citenamefont{Yamashita
  et~al.}(2005)}]{Yamashita:2005:pi-qubit:SFS+SIS}
\bibinfo{author}{\bibfnamefont{T.}~\bibnamefont{Yamashita}},
  \bibnamefont{et~al.}, \bibinfo{journal}{Phys. Rev. Lett.}
  \textbf{\bibinfo{volume}{95}}, \bibinfo{eid}{097001} (\bibinfo{year}{2005}).

\bibitem[{\citenamefont{Yamashita et~al.}(2006)}]{Yamashita:2006:pi-qubit:3JJ}
\bibinfo{author}{\bibfnamefont{T.}~\bibnamefont{Yamashita}},
  \bibnamefont{et~al.}, \bibinfo{journal}{Appl. Phys. Lett.}
  \textbf{\bibinfo{volume}{88}}, \bibinfo{eid}{132501} (\bibinfo{year}{2006}).

\bibitem[{\citenamefont{Weides et~al.}()}]{Weides:2006:SIFS-HiJcPiJJ}
\bibinfo{author}{\bibfnamefont{M.}~\bibnamefont{Weides}}, \bibnamefont{et~al.},
  \eprint{cond-mat/0604097}.

\bibitem[{\citenamefont{Bulaevskii et~al.}(1978)}]{Bulaevskii:0-pi-LJJ}
\bibinfo{author}{\bibfnamefont{L.~N.} \bibnamefont{Bulaevskii}},
  \bibnamefont{et~al.}, \bibinfo{journal}{Solid State Commun.}
  \textbf{\bibinfo{volume}{25}}, \bibinfo{pages}{1053} (\bibinfo{year}{1978}).

\bibitem[{\citenamefont{Xu et~al.}(1995)}]{Xu:SF-shape}
\bibinfo{author}{\bibfnamefont{J.~H.} \bibnamefont{Xu}}, \bibnamefont{et~al.},
  \bibinfo{journal}{Phys. Rev. B} \textbf{\bibinfo{volume}{51}},
  \bibinfo{pages}{11958} (\bibinfo{year}{1995}).

\bibitem[{\citenamefont{Goldobin et~al.}(2002)}]{Goldobin:SF-Shape}
\bibinfo{author}{\bibfnamefont{E.}~\bibnamefont{Goldobin}},
  \bibnamefont{et~al.}, \bibinfo{journal}{Phys. Rev. B}
  \textbf{\bibinfo{volume}{66}}, \bibinfo{pages}{100508(R)}
  (\bibinfo{year}{2002}).

\bibitem[{\citenamefont{Kirtley et~al.}(1997)}]{Kirtley:IcH-PiLJJ}
\bibinfo{author}{\bibfnamefont{J.~R.} \bibnamefont{Kirtley}},
  \bibnamefont{et~al.}, \bibinfo{journal}{Phys. Rev. B}
  \textbf{\bibinfo{volume}{56}}, \bibinfo{pages}{886} (\bibinfo{year}{1997}).

\bibitem[{\citenamefont{Stefanakis}(2002)}]{Stefanakis:ZFS/2}
\bibinfo{author}{\bibfnamefont{N.}~\bibnamefont{Stefanakis}},
  \bibinfo{journal}{Phys. Rev. B} \textbf{\bibinfo{volume}{66}},
  \bibinfo{pages}{214524} (\bibinfo{year}{2002}).

\bibitem[{\citenamefont{Goldobin et~al.}(2003)}]{Goldobin:SF-ReArrange}
\bibinfo{author}{\bibfnamefont{E.}~\bibnamefont{Goldobin}},
  \bibnamefont{et~al.}, \bibinfo{journal}{Phys. Rev. B}
  \textbf{\bibinfo{volume}{67}}, \bibinfo{pages}{224515}
  (\bibinfo{year}{2003}).

\bibitem[{\citenamefont{Zenchuk and Goldobin}(2004)\citenamefont{Zenchuk
  et~al.}}]{Zenchuk:2003:AnalXover}
\bibinfo{author}{\bibfnamefont{A.}~\bibnamefont{Zenchuk}} \bibnamefont{et~al.},
  \bibinfo{journal}{Phys. Rev. B} \textbf{\bibinfo{volume}{69}},
  \bibinfo{pages}{024515} (\bibinfo{year}{2004}).

\bibitem[{\citenamefont{Goldobin
  et~al.}(2004{\natexlab{a}})}]{Goldobin:Art-0-pi}
\bibinfo{author}{\bibfnamefont{E.}~\bibnamefont{Goldobin}},
  \bibnamefont{et~al.}, \bibinfo{journal}{Phys. Rev. Lett.}
  \textbf{\bibinfo{volume}{92}}, \bibinfo{pages}{057005}
  (\bibinfo{year}{2004}{\natexlab{a}}).

\bibitem[{\citenamefont{Susanto et~al.}(2003)}]{Susanto:SF-gamma_c}
\bibinfo{author}{\bibfnamefont{H.}~\bibnamefont{Susanto}},
  \bibnamefont{et~al.}, \bibinfo{journal}{Phys. Rev. B}
  \textbf{\bibinfo{volume}{68}}, \bibinfo{pages}{104501}
  (\bibinfo{year}{2003}).

\bibitem[{\citenamefont{Goldobin et~al.}(2004{\natexlab{b}})}]{Goldobin:F-SF}
\bibinfo{author}{\bibfnamefont{E.}~\bibnamefont{Goldobin}},
  \bibnamefont{et~al.}, \bibinfo{journal}{Phys. Rev. B}
  \textbf{\bibinfo{volume}{70}}, \bibinfo{eid}{094520}
  (\bibinfo{year}{2004}{\natexlab{b}}).

\bibitem[{\citenamefont{Lazarides}(2004)}]{Lazarides:Ic(H):SF-Gen}
\bibinfo{author}{\bibfnamefont{N.}~\bibnamefont{Lazarides}},
  \bibinfo{journal}{Phys. Rev. B} \textbf{\bibinfo{volume}{69}},
  \bibinfo{eid}{212501} (\bibinfo{year}{2004}).

\bibitem[{\citenamefont{Kato and Imada}(1997)\citenamefont{Kato
  et~al.}}]{Kato:1997:QuTunnel0pi0JJ}
\bibinfo{author}{\bibfnamefont{T.}~\bibnamefont{Kato}} \bibnamefont{et~al.},
  \bibinfo{journal}{J. Phys. Soc. Jpn.} \textbf{\bibinfo{volume}{66}},
  \bibinfo{pages}{1445} (\bibinfo{year}{1997}).

\bibitem[{\citenamefont{Koyama et~al.}(2005)}]{Koyama:2005:d-dot:Qu}
\bibinfo{author}{\bibfnamefont{T.}~\bibnamefont{Koyama}}, \bibnamefont{et~al.},
  \bibinfo{journal}{Physica C} \textbf{\bibinfo{volume}{426--431}},
  \bibinfo{pages}{1561} (\bibinfo{year}{2005}).

\bibitem[{\citenamefont{Goldobin et~al.}(2005)}]{Goldobin:2005:MQC-2SFs}
\bibinfo{author}{\bibfnamefont{E.}~\bibnamefont{Goldobin}},
  \bibnamefont{et~al.}, \bibinfo{journal}{Phys. Rev. B}
  \textbf{\bibinfo{volume}{72}}, \bibinfo{eid}{054527} (\bibinfo{year}{2005}).

\bibitem[{\citenamefont{Van~Harlingen}(1995)}]{VanHarlingen:1995:Review}
\bibinfo{author}{\bibfnamefont{D.~J.} \bibnamefont{Van~Harlingen}},
  \bibinfo{journal}{Rev. Mod. Phys.} \textbf{\bibinfo{volume}{67}},
  \bibinfo{pages}{515} (\bibinfo{year}{1995}).

\bibitem[{\citenamefont{Tsuei and Kirtley}(2000)\citenamefont{Tsuei
  et~al.}}]{Tsuei:Review}
\bibinfo{author}{\bibfnamefont{C.~C.} \bibnamefont{Tsuei}}
  \bibnamefont{et~al.}, \bibinfo{journal}{Rev. Mod. Phys.}
  \textbf{\bibinfo{volume}{72}}, \bibinfo{pages}{969} (\bibinfo{year}{2000}).

\bibitem[{\citenamefont{Chesca
  et~al.}(2002)}]{Chesca:2002:YBCO4Crystal-pi-SQUID}
\bibinfo{author}{\bibfnamefont{B.}~\bibnamefont{Chesca}}, \bibnamefont{et~al.},
  \bibinfo{journal}{Phys. Rev. Lett.} \textbf{\bibinfo{volume}{88}},
  \bibinfo{pages}{177003} (\bibinfo{year}{2002}).

\bibitem[{\citenamefont{Smilde et~al.}(2002)}]{Smilde:ZigzagPRL}
\bibinfo{author}{\bibfnamefont{H.-J.~H.} \bibnamefont{Smilde}},
  \bibnamefont{et~al.}, \bibinfo{journal}{Phys. Rev. Lett.}
  \textbf{\bibinfo{volume}{88}}, \bibinfo{pages}{057004}
  (\bibinfo{year}{2002}).

\bibitem[{\citenamefont{Ariando et~al.}(2005)}]{Ariando:Zigzag:NCCO}
\bibinfo{author}{\bibnamefont{Ariando}}, \bibnamefont{et~al.},
  \bibinfo{journal}{Phys. Rev. Lett.} \textbf{\bibinfo{volume}{94}},
  \bibinfo{eid}{167001} (\bibinfo{year}{2005}).

\bibitem[{\citenamefont{Kirtley et~al.}(1999)}]{Kirtley:SF:T-dep}
\bibinfo{author}{\bibfnamefont{J.~R.} \bibnamefont{Kirtley}},
  \bibnamefont{et~al.}, \bibinfo{journal}{Science}
  \textbf{\bibinfo{volume}{285}}, \bibinfo{pages}{1373} (\bibinfo{year}{1999}).

\bibitem[{\citenamefont{Sugimoto et~al.}(2002)}]{Sugimoto:TriCrystal:SF}
\bibinfo{author}{\bibfnamefont{A.}~\bibnamefont{Sugimoto}},
  \bibnamefont{et~al.}, \bibinfo{journal}{Physica C}
  \textbf{\bibinfo{volume}{367}}, \bibinfo{pages}{28} (\bibinfo{year}{2002}).

\bibitem[{\citenamefont{Hilgenkamp et~al.}(2003)}]{Hilgenkamp:zigzag:SF}
\bibinfo{author}{\bibfnamefont{H.}~\bibnamefont{Hilgenkamp}},
  \bibnamefont{et~al.}, \bibinfo{journal}{Nature (London)}
  \textbf{\bibinfo{volume}{422}}, \bibinfo{pages}{50} (\bibinfo{year}{2003}).

\bibitem[{\citenamefont{Kirtley et~al.}(2005)}]{Kirtley:2005:AFM-SF}
\bibinfo{author}{\bibfnamefont{J.~R.} \bibnamefont{Kirtley}},
  \bibnamefont{et~al.}, \bibinfo{journal}{Phys. Rev. B}
  \textbf{\bibinfo{volume}{72}}, \bibinfo{eid}{214521} (\bibinfo{year}{2005}).

\bibitem[{\citenamefont{Della~Rocca
  et~al.}(2005)}]{DellaRocca:2005:0-pi-SFS:SF}
\bibinfo{author}{\bibfnamefont{M.~L.} \bibnamefont{Della~Rocca}},
  \bibnamefont{et~al.}, \bibinfo{journal}{Phys. Rev. Lett.}
  \textbf{\bibinfo{volume}{94}}, \bibinfo{eid}{197003} (\bibinfo{year}{2005}).

\bibitem[{\citenamefont{Frolov et~al.}(2006)}]{Frolov:2006:SFS-0-pi}
\bibinfo{author}{\bibfnamefont{S.~M.} \bibnamefont{Frolov}},
  \bibnamefont{et~al.}, \bibinfo{journal}{Phys. Rev. B}
  \textbf{\bibinfo{volume}{74}}, \bibinfo{eid}{020503} (\bibinfo{year}{2006}).

\bibitem[{\citenamefont{Weides et~al.}(2006)}]{Weides:SIFS-Tech}
\bibinfo{author}{\bibfnamefont{M.}~\bibnamefont{Weides}}, \bibnamefont{et~al.},
  \bibinfo{journal}{Physica C} \textbf{\bibinfo{volume}{437--438}},
  \bibinfo{pages}{349} (\bibinfo{year}{2006}).

\bibitem[{\citenamefont{Vohl et~al.}(1991)}]{Vohl:1991:WedgeF-layer}
\bibinfo{author}{\bibfnamefont{M.}~\bibnamefont{Vohl}}, \bibnamefont{et~al.},
  \bibinfo{journal}{J. Magn. Magn. Mat.} \textbf{\bibinfo{volume}{93}},
  \bibinfo{pages}{403} (\bibinfo{year}{1991}).

\bibitem[{\citenamefont{Kim et~al.}(2005)}]{Kim:2005:Proxy:Nb-Au-CoFe}
\bibinfo{author}{\bibfnamefont{J.}~\bibnamefont{Kim}}, \bibnamefont{et~al.},
  \bibinfo{journal}{Phys. Rev. B} \textbf{\bibinfo{volume}{71}},
  \bibinfo{eid}{214519} (\bibinfo{year}{2005}).

\bibitem[{\citenamefont{Zdravkov
  et~al.}(2006)}]{Zdravkov:2006:SC:Nb-CuNi-bilayers}
\bibinfo{author}{\bibfnamefont{V.}~\bibnamefont{Zdravkov}},
  \bibnamefont{et~al.}, \bibinfo{journal}{Phys. Rev. Lett.}
  \textbf{\bibinfo{volume}{97}}, \bibinfo{eid}{057004} (\bibinfo{year}{2006}).

\bibitem[{\citenamefont{Gurvitch et~al.}(1983)}]{Gurvitch:1983:NbTech}
\bibinfo{author}{\bibfnamefont{M.}~\bibnamefont{Gurvitch}},
  \bibnamefont{et~al.}, \bibinfo{journal}{Appl. Phys. Lett.}
  \textbf{\bibinfo{volume}{42}}, \bibinfo{pages}{472} (\bibinfo{year}{1983}).

\end{thebibliography}

\end{document}